# Microwave magnetoabsorption in $R_{0.6}Sr_{0.4}MnO_3$ (R = Pr and Nd)

A. Chanda, U. Chaudhuri and R. Mahendiran[1]

Physics Department, 2 Science Drive 3, National University of Singapore,

Singapore-117551, Republic of Singapore

**Abstract**

We report microwave magnetoabsorption ($\Delta P$) at room temperature in $R_{0.6}Sr_{0.4}MnO_3$ (R = Pr and Nd) samples. $\Delta P$ as a function of *dc* magnetic field (-2.5 kOe ≤ $H_{dc}$ ≤ 2.5 kOe) was measured for a broad frequency range ($f$ = 0.1 - 4 GHz) of microwave magnetic field using a vector network analyzer and a copper strip coil which encloses one of the above samples. As external dc magnetic field decreased from the maximum field, $\Delta P$ initially increases and shows a maximum for a critical field and then decreases as the field approaches zero. The critical field increases with increasing frequency of the microwave signal. Line shape analysis of the obtained spectra suggest that the observed features in $\Delta P$ are caused by ferromagnetic resonance (FMR) in $R$ = Pr and electron paramagnetic resonance (ESR) in $R$ = Nd samples, which were confirmed using a co-planar waveguide based broadband magnetic resonance spectrometer.

---

[1] Author for correspondence(phyrm@nus.edu.sg)

## 1. Introduction

Doped manganites of the formula $R_{1-x}A_x MnO_3$ where $R = La^{3+}$, $Pr^{3+}$, $Nd^{3+}$ etc., and $A = Sr^{2+}$, $Ca^{2+}$ etc., are well known for the colossal magnetoresistance (CMR) effect -a dramatic decrease of electrical resistance under external magnetic fields exhibited by them. [1] An interesting and yet relatively less explored property of these oxides is the magnetic field dependence of microwave absorption (MWA). Available reports indicate that microwave absorption in some members of this oxide family is more sensitive than dc electrical resistance to external magnetic fields, particularly in the vicinity of paramagnetic to ferromagnetic transition. For example, Owens[2] found 50 % change in microwave surface resistance in polycrystalline $La_{0.7}Sr_{0.3}MnO_3$ in a small magnetic field of $H = 600$ Oe and Srinivasu et al. [3] reported still higher change (~80%) for the same field strength in $La_{0.7}Ba_{0.3}MnO_3$ powder but Tyagi et al. [4] found smaller magnetoabsorption (~20% for $H = 10$ kOe) in $La_{0.7}Ba_{0.3}MnO_3$ thin film. A maximum MWA occurs when the resonance frequency of the impinging microwave magnetic field coincides with precessional frequency of the magnetization. MWA can also occur near zero-field due to domain magnetization process or magnetoresistance in a ferromagnetic sample which causes low-field non-resonant microwave absorption.[5,6] Magnetoabsorption was used to detect magnetization avalanche in a phase-separated $La_{0.225}Pr_{0.4}Ca_{0.375}MnO_3$.[7] In majority of the available reports, a commercial electron spin resonance (ESR) spectrometer was used to record the MWA spectra while sweeping an external dc magnetic field at a single frequency (~9.8 GHz) dictated by the natural resonance frequency of the resonant cavity.[8,9,10,11] There are few reports of frequency swept microwave absorption in zero external magnetic field in $La_{1-x}Sr_xMnO_3$ ($x = 0.2$-$0.33$) samples using a vector network analyser (VNA) and specially designed transmission line[12] or shorted coaxial cables provided by a manufacturer of VNA. [13,14,15,16,17] However, broadband microwave absorption in the presence of a magnetic field was not studied in those work. Also, there is no previous report on the broadband microwave absorption in the paramagnetic state of manganites.

Here, we report microwave magnetoabsorption (MWMA) in ferromagnetic ($R$ = Pr) and paramagnetic ($R$ = Nd) samples of the composition $R_{0.6}Sr_{0.4}MnO_3$ over a broad frequency range (0.1 - 4 GHz) by making use of a vector network analyzer (VNA) and a copper strip coil. The strip coil serves as a trans-receiver to radiate microwave field and sense the high-frequency magnetic response of the material. Novelties of the present work are: (1) Our broadband FMR/ESR technique makes use of a copper strip-coil instead of a microfabricated coplanar-waveguide (CPW) or microstrip, (2) We observe a frequency tunable peak in the field dependence of microwave absorption, which we attribute to FMR in the Pr based manganite and ESR in the Nd based manganite, (3) Our interpretations are supported by the detection of FMR and ESR signals using a CPW based commercial spectrometer which makes uses of the lock-in technique and measures the microwave absorption as a field derivation of power absorption ($dP/dH$).

## 2. Experimental details

Polycrystalline samples of $R_{0.6}Sr_{0.4}MnO_3$ ($R$ = La, Pr and Nd) were prepared by conventional solid-state reaction method. Magnetization of these samples were measured using a vibrating sample magnetometer incorporated to a physical property measurement system (Quantum Design Inc., USA). For the MWMA measurement, these samples were cut into rectangular geometry with dimensions of 3 mm length x 3 mm width x 2 mm thickness. A cuboidal shaped strip coil of similar dimensions was made from a copper sheet of 0.2 mm thickness. A schematic illustration of our experimental setup is shown in the Fig. 1. The sample was firmly fit into the strip coil and the inner surface of coil was also covered with a thin layer of a Kapton tape to insulate the sample from touching the sample. Two ends of the coil were connected to a VNA (Agilent model N5230A) using SMA connectors and radio frequency (RF) cables. The Agilent model N5230A is a single port VNA which supplies microwave power to the load (copper strip) and measures reflected power using the same port. The output power of VNA was maintained at a constant power mode (10 dBm). A maximum power will be delivered to the strip coil if its impedance is matched to 50 $\Omega$[18]. However, our copper strip coil is not impedance matched. Impedance matching is not essential in our case as we are interested in relative change of MWA without and with an external dc magnetic field. An electromagnet was used to sweep dc magnetic field ($H_{dc}$). Magnetic field dependence of microwave power absorption was measured in terms of $S_{11}$ scattering parameter of the electromagnetic wave reflected from the load upon change in microwave property of the sample while sweeping the dc magnetic field. The VNA measures the $S_{11}$ parameter in terms of decibels (dB) or dBm. The SMA connector which links the strip coil and the RF cable has traces of magnetic impurities which give non-negligible contribution to $\Delta P$ spectra of the paramagnetic sample. To eliminate the contribution from the SMA connector, data were also taken with the empty coil for each frequency and subtracted from the data measured with the sample inside the coil.

The field derivative of microwave power absorption ($dP/dH$) as a function of magnetic field at room temperature was measured using a commercial broadband ferromagnetic resonance spectrometer (NanOsc Cryo FMR$^{TM}$, Quantum Design Inc., USA) integrated to a Physical Property Measurement System (Quantum Design Inc., USA).

## 3. Results

The main panel of Fig. 2 shows the temperature dependence of magnetization, $M(T)$ of $R_{0.6}Sr_{0.4}MnO_3$ ($R$ = Pr and Nd) samples measured in a dc magnetic field of $H$ = 1 kOe. The ferromagnetic Curie temperature ($T_C$) determined from the minimum of $dM/dT$ curves are $T_C$ = 305 and 273 K for the samples $R$ = Pr and Nd, respectively. Thus, the sample $R$ = Pr is ferromagnetic at room temperature whereas the sample $R$ = Nd is a room temperature paramagnet. Magnetic field dependence of magnetization ($M$) at room temperature is shown in the inset for both the samples. $M(H)$ of $R$ = Pr shows typical behavior of a soft ferromagnet- a rapid increase in low magnetic fields followed by approach to saturation at higher fields. On the other hand, $M$ increases linearly with $H$ for $R$ = Nd sample which confirms paramagnetism in this sample at room temperature.

We compare the dc magnetic field ($H_{dc}$) dependence of the change in microwave power absorption ($\Delta P$) at room temperature for samples $R$ = Pr and Nd in the Figs. 3(a) and (b), respectively, for three representative frequencies of MW electromagnetic field ($f$ = 0.1, 1 and 4 GHz). $P$ value at the highest field is taken as the reference. For $R$ = Pr and $f$ = 0.1 GHz, $\Delta P$ vs $H_{dc}$ curve exhibits a single peak at the origin $H_{dc}$ = 0. It implies that the MW power absorption decreases as the magnetic field is increased from zero field. When $f$ = 1 GHz, $\Delta P$ shows broad peaks around $H_{dc} = \pm H_p = \pm 178$ Oe on either sides of the origin and a dip at $H_{dc}$ = 0. At $f$ = 4 GHz, the magnitude of the peak is enhanced and $H_p$ is shifted to a higher magnetic field ($\pm H_p = \pm 772$ Oe). The $\Delta P$ vs $H_{dc}$ curve for $R$ = Nd does not show a peak at the origin for $f$ = 0.1 GHz and the signal is weak compared to the ferromagnetic sample. A double peak structure is visible at $H_{dc} = \pm H_c$ for $f$ = 1 GHz and 4 GHz and the peak position shifts to higher magnetic fields as the frequency increases. However, $\Delta P$ is nearly field independent between $+H_c$ and $-H_c$ for the Nd sample in contrast to the Pr sample for which $\Delta P$ smoothly increases as the dc field is increased from 0 Oe to $+H_p$ or $-H_p$. It should be pointed out that $H_c > H_p$ for a fixed frequency. For $f = 4$ GHz, $H_c = 1.8H_p$. Also, the shift in the peak position ($\Delta H_c = 925$ Oe) for the Nd sample is higher than that of Pr sample ($\Delta H_p = 594$ Oe) for $\Delta f = 3$ GHz. Thus, the anomalous features observed in MWMA are distinct in Pr and Nd samples.

We have plotted the three dimensional graph of $\Delta P$ as a function of frequency of the electromagnetic field ($f = 0.1 - 4$ GHz) and dc magnetic field ($H_{dc}$) for $R = $ Pr and Nd in Figs. 3 (c) and (d), respectively. It is obvious that broad peaks at $H_{dc} = \pm H_p$ for the Pr sample and sharp peaks at $H_{dc} = \pm H_c$ for the Nd sample move apart from each other towards higher $H_{dc}$ values with increasing frequency of the microwave electromagnetic field. The peaks at $H_{dc} = \pm H_c$ for the $R = $ Nd sample are much sharper and positioned at higher field values in contrast to broad peaks at $H_{dc} = \pm H_p$ for the $R = $ Pr sample.

## 4. Discussion

What could be the origin of the observed anomalies in microwave power absorption in the samples studied here? It is known that ferromagnetism in manganites originates from the double exchange interaction between $Mn^{3+}(t_{2g}^3 e_g^1)$ and $Mn^{4+}(t_{2g}^3 e_g^0)$ ions. Both localized $t_{2g}^3$ core spins and itinerant $e_g$ electron's spin of Mn ions collectively contribute to overall magnetization of the sample. In our measurement configuration, propagation of microwave (MW) in the strip coil creates a MW magnetic field ($H_{mw}$) along the axis of coil and hence through the length of the sample. The microwave magnetic field $H_{mw}$ and the applied dc magnetic field $H_{dc}$ are orthogonal to each other as like the field configuration in a conventional ESR spectrometer. Hence, the magnetization undergoes a precessional motion around the effective magnetic field ($H_{eff}$) which is the combination of anisotropy field ($H_k$) and externally applied dc magnetic field $H_{dc}$. A resonant absorption occurs when the frequency of the microwave magnetic field matches with the frequency of magnetization precision. This phenomenon is known as electrons spin resonance (ESR) in the case of a paramagnetic sample and ferromagnetic resonance (FMR) in the case of a ferromagnetic sample. Since the Nd sample is paramagnetic at room temperature, the sharp peak absorbed in MWMA is caused ESR. A collective precession of the exchange coupled $Mn^{4+}$ (S= 3/2) ions via $Mn^{3+}$ (S = 2) spin system gives rise to ESR in paramagnetic manganites.[19] In an external dc magnetic field, degenerate $t_{2g}$ and $e_g$ levels of the Mn ions Zeeman split into spin up and spin down energy levels. If the sample is subject to microwave magnetic field in the transverse direction to the dc magnetic field, resonance occurs when the energy gap between the Zeeman-split spin-up and spin-down levels matches with the energy of the microwave magnetic field. Under the resonance condition, the paramagnetic sample absorbs maximum power from the microwave field and the $t_{2g}$ and $e_g$ spins undergo a spin-flip transition from lower to higher energy levels. When resonance occurs, magnetic field ($H_{dc}$) dependence of the out of phase component of permeability ($\mu''$) shows a maximum at $H_{dc} = H_{res}$ where $H_{res}$ is called the resonance field. The power absorbed (*P*) by a sample is related to $\mu''$ through the relation $P = \frac{1}{2}V\mu''\omega h_{mw}^2$, where, $h_{mw}$ is the amplitude of the microwave magnetic field and *V* is the sample volume through which MW field penetrates.[20] Hence, while sweeping the external magnetic field, *P(H)* goes through a peak value at $H_{res}$. Thus, the observed features in microwave magnetoabsorption can be ascribed to FMR for *R* = Pr and EPR for *R* = Nd samples. The single peak feature in the *ΔP* vs *H* curves at $H_{dc}$ = 0 for frequencies below 0.5 GHz for the ferromagnetic sample *R* = Pr is a non-resonant absorption feature caused by magnetoresistance and domain rotation process.[5,6]

The magnetic field dependence of microwave power absorption, $P(H_{dc})$, usually follows the Lorentzian function described by,[21]

$$P(H_{dc}) = P_{max} \frac{\left(\frac{\Delta H}{2}\right)^2}{(H_{dc}-H_{res})^2+\left(\frac{\Delta H}{2}\right)^2} \qquad (1)$$

where, $P_{max}$ is the maximum power absorption at $H_{dc} = H_{res}$ and $\Delta H$ is the linewidth (full line width at half maximum (FWHM)). However, in a conducting ferromagnet, penetration of MW magnetic field is limited to the skin depth ($\delta$) which is controlled by the frequency dependence of the permeability ($\mu$) of the sample through the relation: $\delta = \sqrt{\frac{2\rho}{2\pi f \mu_0 \mu}}$, where $\rho$ is the dc resistivity and $\mu_0$ is the free space permeability. In the limit of strong skin effect, the power absorption is proportional $\sqrt{|\mu|+\mu''}$ where $\mu = \mu' - i\mu''$ and dispersive ($\mu'$) component mixes with the absorption component ($\mu''$)[20]. While the absorptive component is described by a symmetric Lorentzian function (as shown in the Eqn. 1) the dispersive component follows the antisymmetric Lorentzian function. Hence, the resultant power absorption spectrum for a metallic ferromagnet can be described by a linear combination of symmetric and antisymmetric Lorentzian functions as:[22]

$$P(H_{dc}) = P_{sym} \frac{\left(\frac{\Delta H}{2}\right)^2}{(H_{dc}-H_{res})^2+\left(\frac{\Delta H}{2}\right)^2} + P_{asym} \frac{\frac{\Delta H}{2}(H_{dc}-H_{res})}{(H_{dc}-H_{res})^2+\left(\frac{\Delta H}{2}\right)^2} + P_0 \qquad (2)$$

where, $P_{sym}$ and $P_{asym}$ are the coefficients of symmetric and antisymmetric Lorentzian functions, respectively and $P_0$ is a constant offset parameter. Although Eqn. 1 and Eqn. 2 are valid for narrow resonance line shapes they have been used to estimate the linewidths qualitatively.

At $T = 300$ K, dc resistivity of our samples are r = 54.7 mW cm and 76.9 mW cm for R = Pr and Nd samples, respectively. Assuming $\mu = 1$, the nonmagnetic ($\mu = 1$) skin depth ($\delta$) for the R = Pr and Nd samples are $\delta = 186$ and 221 µm, respectively for $f = 4$ GHz, which are larger than the average grain size (~3 µm) of the samples but smaller than dimensions (2 mm thick, 3 mm in width ) of the sample. Hence, the influence of eddy current need to be considered while analyzing the resonance line shapes. Recently, Flovik et al. [23] showed that dispersive line shape arising from the out-of-phase driving fields induced by eddy currents can also be described by an antisymmetric Lorentzian function. Thus, the second term in Eqn. 2 accounts for the contributions from both the dispersive component of permeability as well as the eddy current effects.

Generally, for a paramagnetic sample, the magnetic field dependence of MW power absorption, $P(H_{dc})$ is described by the Dysonian function as:[24,25]

$$P(H_{dc}) = I_0 \frac{\Delta H + \beta(H_{dc}-H_{res})}{4(H_{dc}-H_{res})^2+(\Delta H)^2} \qquad (3)$$

where, $I_0$ is the signal intensity and $\beta$ ($0 \leq \beta \leq 1$) is the asymmetry parameter. The asymmetry parameter is the ratio of the dispersive component to the absorptive component of the resultant signal. Note that the Eqn. (3) is valid for a narrow resonance line ($\Delta H << H_{res}$). On the other hand, the field dependence of the resultant MW power absorption for broad resonance line shapes ($\Delta H >> H_{res}$) are described by the following expression:[26]

$$P(H_{dc}) = I_0 \left[\frac{\Delta H + \beta(H_{dc}-H_{res})}{4(H_{dc}-H_{res})^2+(\Delta H)^2} + \frac{\Delta H + \beta(H_{dc}+H_{res})}{4(H_{dc}+H_{res})^2+(\Delta H)^2}\right] \qquad (4)$$

Since for the $R$ = Nd sample, $\Delta H << H_{res}$ ($H_{res}$ ~ 1.35 kOe and $\Delta H < 0.5$ kOe at $f = 4$ GHz), the line shapes can be described by Eqn. 3. Multiplying both the numerator and the denominator of the Eqn. (3) by $\Delta H$, we get,

$$P(H_{dc}) = \frac{I_0}{\Delta H} \cdot \frac{(\Delta H)^2 + \beta \Delta H(H_{dc}-H_{res})}{4(H_{dc}-H_{res})^2+(\Delta H)^2} = \frac{I_0}{\Delta H} \cdot \frac{\left(\frac{\Delta H}{2}\right)^2}{(H_{dc}-H_{res})^2+\left(\frac{\Delta H}{2}\right)^2} + \frac{\beta I_0}{2\Delta H} \cdot \frac{\frac{\Delta H}{2}(H_{dc}-H_{res})}{(H_{dc}-H_{res})^2+\left(\frac{\Delta H}{2}\right)^2} \qquad (5)$$

Thus, the expression of $P(H_{dc})$ described by Eqn. 5 is identical to the Eqn. (2) with $P_{sym} = I_0/\Delta H$ and $P_{asym} = \beta I_0/2\Delta H$. Hence, the $\Delta P(H_{dc})$ line shapes for both ferromagnetic and paramagnetic resonances can be described by Eqn. 2. We fitted the $\Delta P(H_{dc})$ line shape at frequencies higher than 2 GHz using the Eqn. (2) for both the samples. Fig. 4(a) shows the fit of the $\Delta P$ vs $H_{dc}$ line shapes using the above equation at selected frequencies for the ferromagnetic sample $R$ = Pr. The main panel of Fig. 4(b) shows the plot of frequency ($f$) vs. resonance field ($H_{res}$) for the $R$ = Pr sample. The $f$ vs. $H_{res}$ curve for this sample follows the Kittel's equation for ferromagnetic resonance,[26]

$$f = \frac{1}{2\pi}\left[\{\gamma H_{dc} + (N_x - N_z + N_x^a)\omega_0\}\{\gamma H_{dc} + (N_y - N_z + N_y^a)\omega_0\}\right]^{1/2} \qquad (6)$$

where, $\omega_0 = 4\pi\gamma M_S$, $N_x$, $N_y$ and $N_z$ are the demagnetization factors for the sample geometry, and $N_x^a$ and $N_y^a$ are the demagnetization factors due to magnetocrystalline anisotropy and $M_S$ is the saturation magnetization. $\gamma$ is the gyromagnetic ratio given by $\gamma = g\mu_B/\hbar$, where "$g$" is the "Lande' $g$-factor". The demagnetization factors can be determined assuming ellipsoidal geometry of the sample.[27] For our sample dimensions, the values of $N_x$, $N_y$ and $N_z$ are 0.272, 0.272, and 0.456, respectively. The magnetocrystalline anisotropy is considered to be in the plane of the sample (along the $x$-direction) but perpendicular to the applied dc magnetic field. We have also assumed that the out of plane component of the magnetocrystalline anisotropy is negligible. Hence, $N_x^a = H_k/M_S$, $N_y^a = 0$, where, $H_k$ is the anisotropy field. The value of $\gamma/2\pi$ for the Pr sample extracted from the Kittel fit described by Eqn. (6) is 2.82 MHz/Oe, which is close to the free electron value ($\gamma/2\pi$ =2.8 MHz/Oe). We also obtained $4\pi M_s$ = 1417 ± 100 Oe and $H_k$ = 232 ± 10 Oe, respectively. The value of $H_k$ obtained from the magnetoabsorption data for the sample $R$ = Pr closely matches with $H_k$ = 206 ± 15 Oe, estimated by fitting the $M(H)$ isotherm of this sample (see inset of Fig. 2(b)) with the law of approach to saturation model (LAS) which is usually expressed as:[28]

$$M = M_s\left(1 - \frac{a}{H} - \frac{b}{H^2}\right) + \chi H \qquad (7)$$

where, the coefficients "a" and "b" are connected to micro-stress and the first order magnetocrystalline anisotropy coefficient (K), respectively. The anisotropy field $H_k$ can be calculated using the relation: $H_k = 2K/\mu_0 M_s$. The value of $M_S$ obtained from the LAS fit is 28 emu/g. Using the value of mass density (estimated by the Archimedes principle), $\rho_m = 3.25 \times 10^3$ kg/m$^3$, the calculated value of $4\pi M_s$ for the R = Pr sample is 1143.5 Oe, which is slightly lower than the value of $4\pi M_s = 1417 \pm 100$ Oe obtained from the Kittel fit. The discrepancy between these two $M_S$ values could arise from the fact that the measurement is done at 300 K, which is closer to the ferromagnetic Curie temperature ($T_C$ = 305 K) and the magnetization is not saturated in a field of 2.5 kOe at room temperature. The inset of Fig. 4(b) shows variation of linewidth $\Delta H$ with frequency and it shows non-linear frequency dependence in the measured frequency range. "The large line width seen in the ferromagnetic sample needs to be understood. The ferromagnetic Curie temperature of the Pr sample ($T_C$ = 305 K) is not far from the temperature of the measurement. Other than intrinsic Gilbert damping mechanism, extrinsic contributions such as magnetic inhomogeneity within grains, porosity, two magnon scattering and eddy current in the sample could lead to broadening of the line width.[29,30] Since our samples are polycrystalline samples, it is highly probable that the extrinsic effects dominate and lead to a nonlinear scaling of the line width with frequency. Measurements as a function of temperature is needed to understand the exact mechanism of line broadening."

Fig. 5(a) shows the fitting of $\Delta P$ ($H_{dc}$) line shapes at selected frequencies for the paramagnetic sample R = Nd using the Eqn. (2). The frequency dependence of $\Delta H$ extracted from the fitting is shown in the inset of Fig. 5(b) which also shows a non-linear frequency dependence. The values of $\Delta H$ for this sample are clearly much smaller than the ferromagnetic sample R = Pr. The main panel of Fig. 5(b) clearly indicates that $H_{res}$ increases linearly with $f$ and thus follows the condition for electron paramagnetic resonance, i.e., $f = (\gamma/2\pi) H_{res}$. A linear fit to the $f$ vs. $H_{res}$ curve yields $\gamma/2\pi = 2.87$ MHz/Oe which is close to the free electron value.

In order to confirm our observations of FMR and ESR in the samples $R$ = Pr and Nd, magnetic resonance measurements were also performed on both these samples using a commercial broadband FMR spectrometer. This measurement makes use of a coplanar waveguide (CPW) based technique and records the field derivative of MW power absorption ($dP/dH$) by the sample placed on the CPW when excited by a MW magnetic field generated by the MW current flowing through the CPW. The left y-scales of Figs. 6(a) and (b) depict $dP/dH$ as a function of dc magnetic field ($H_{dc}$) at a fixed frequency of MW excitation $f = 4$ GHz for the samples $R$ = Pr and Nd, respectively. Numerical integration was performed on the $dP/dH$ spectra to estimate the corresponding change in MW power absorption ($\Delta P$) as a function of $H_{dc}$ (see right y-scales of Figs. 6(a) and (b)). It is clear that the line shapes of $\Delta P$ line obtained from the Cryo-FMR technique for both the samples are similar to that obtained from strip-coil-VNA technique at the same frequency. Furthermore, the positions of the resonance fields ($H_{res}$) at $f = 4$ GHz obtained from both the techniques closely match with each other for both $R$ = Pr and Nd samples which validates the observations of FMR and ESR in these samples using our broadband detection method.

## 5. Summary

In summary, we have investigated microwave absorption of ferromagnetic ($R$ = Pr) and paramagnetic ($R$ = Nd) samples in $R_{0.6}Sr_{0.4}MnO_3$ as a function of dc magnetic field over a broad frequency range (0.1 - 4 GHz) using a vector network analyzer and a copper strip coil at room temperature. Our microwave magnetoabsorption results show features of FMR for the sample $R$ = Pr and ESR for the sample $R$ = Nd, which were further confirmed by a coplanar waveguide based broadband FMR spectrometer. By performing the line shape analysis, we have extracted the gyromagnetic ratio, saturation magnetization and anisotropy field for the ferromagnetic sample.

**Acknowledgements:** R. M. thanks the Ministry of Education for supporting this work (Grant numbers: R144-000-373-112 and R144-000-381-112).

**Figure Captions**

**FIG. 1.** A schematic representation of the strip-coil based microwave power absorption setup.

**FIG. 2**. Temperature dependence of magnetization (*M*) of the samples $R$ = Pr and Nd in $R_{0.6}Sr_{0.4}MnO_3$ series in a magnetic field of $H_{dc}$ = 1 kOe. Inset: *M-H* loop at 300 K for the same samples along with the Fit using the Law of Approach to Saturation Model for the sample $R$ = Pr.

**FIG. 3.** Magnetic field dependence of the change in microwave power absorption (*ΔP*) at room temperature for the samples (a) $R$ = Pr and (b) $R$ = Nd at three selected frequencies (*f* = 0.1, 3 and 4 GHz) at room temperature. Bottom: 3D plots of *ΔP* vs $H_{dc}$ curves at different frequencies (*f* = 0.1 - 4 GHz) for samples (c) $R$ = Pr and (d) $R$ = Nd.

**FIG. 4**(a) Lorentzian fit of the *ΔP* line shapes for the sample $R$ = Pr at selected frequencies between *f* = 2 - 4 GHz, (b) Main panel: *f-$H_{res}$* curve obtained from the *ΔP* line shape analysis for the sample $R$ = Pr with the Kittel fit and inset shows *f* dependence of the line width *ΔH*.

**FIG. 5**(a) Lorentzian fit of the *ΔP* line shape for $R$ = Nd sample at selected frequencies between *f* = 2 - 4 GHz, (b) Main panel: liner fit to the *f-$H_{res}$* curve obtained from the *ΔP* line shape analysis and inset shows *f* dependence of the line width *ΔH*.

**FIG. 6** Magnetic field dependence of the field derivative of the microwave power absorption (*dP/dH*) at *f* = 4 GHz [left *y*-scale] measured using broadband FMR spectrometer and corresponding change in power absorption (*ΔP*) estimated by integrating the *dP/dH* data over the measured field range (right-*y* scale) for the samples (a) $R$= Pr and (b) $R$ = Nd.

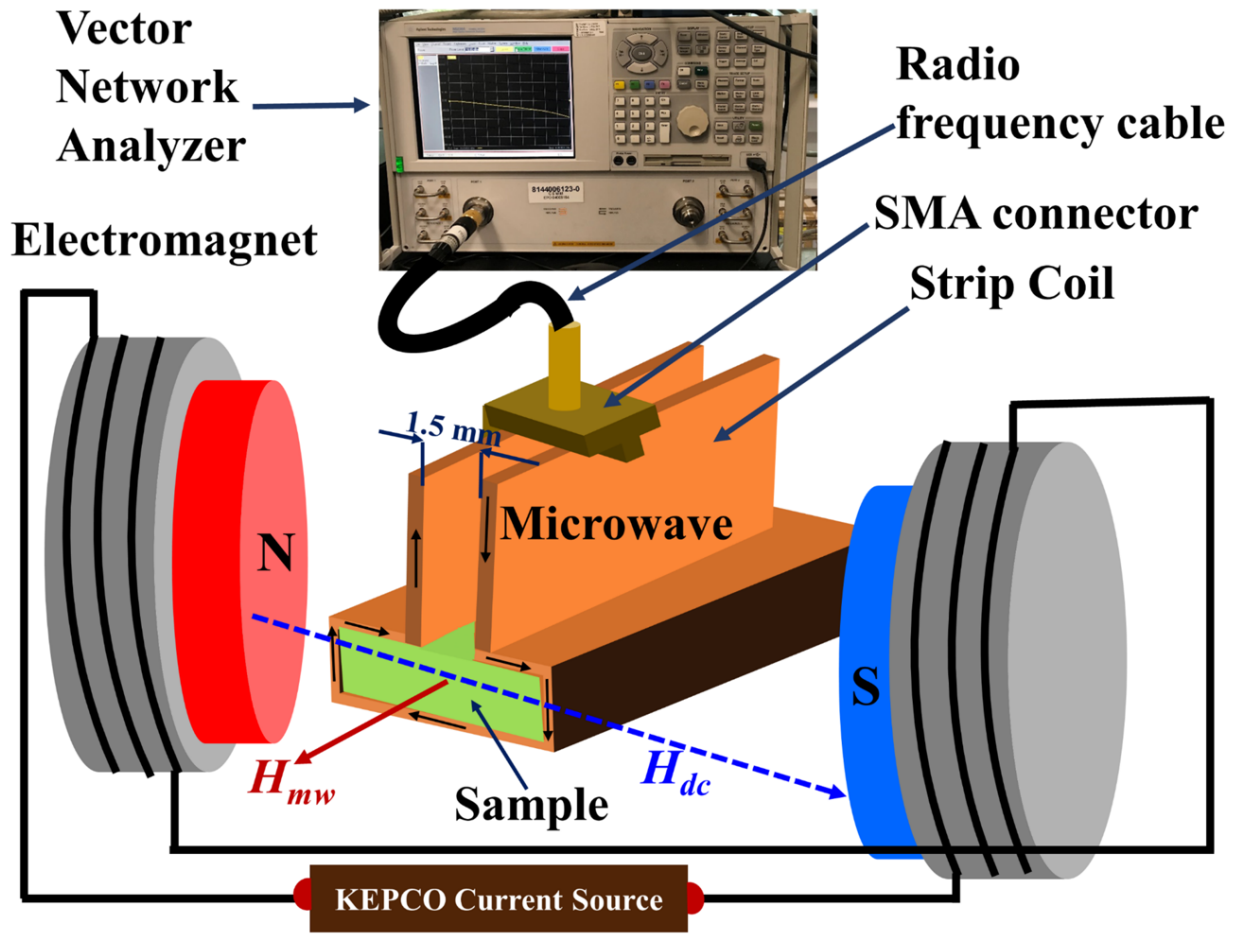

**FIG. 1.** A. Chanda et al.

**FIG. 1.** A schematic representation of the strip-coil based microwave power absorption setup.

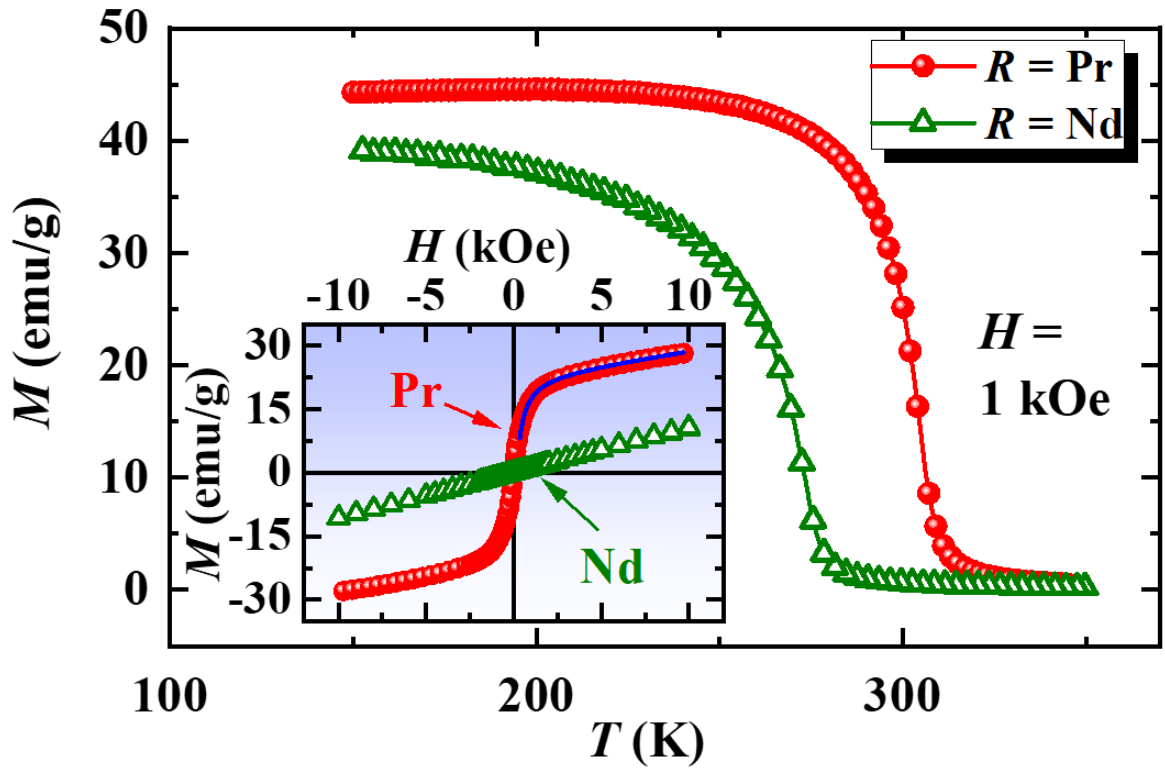

Fig. 2. A. Chanda et al.

**FIG. 2**. Temperature dependence of magnetization (*M*) of the samples *R* = Pr and Nd in $R_{0.6}Sr_{0.4}MnO_3$ series in a magnetic field of $H_{dc}$ = 1 kOe. Inset: *M-H* loop at 300 K for the same samples along with the Fit using the Law of Approach to Saturation Model for the sample *R* = Pr.

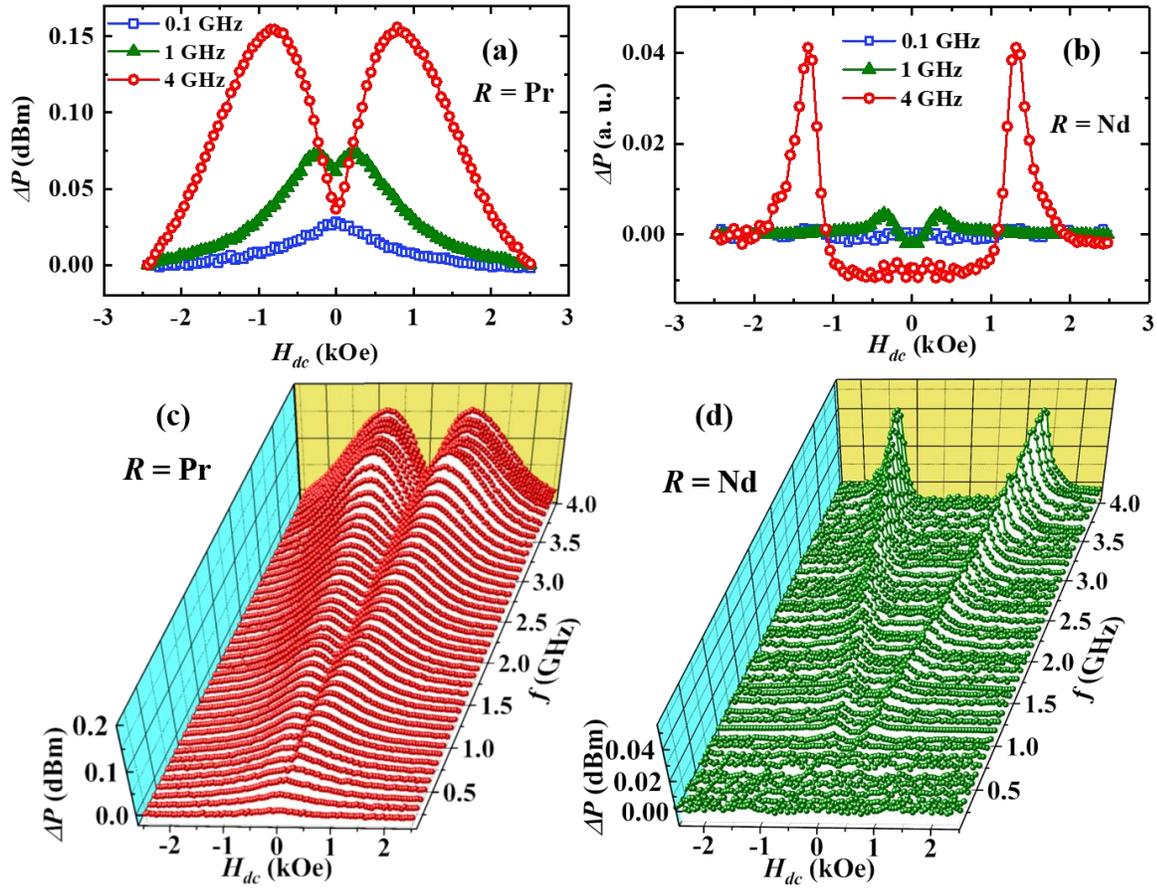

**FIG. 3.** Magnetic field dependence of the change in microwave power absorption ($\Delta P$) at room temperature for the samples (a) $R$ = Pr and (b) $R$ = Nd at three selected frequencies ($f$ = 0.1, 3 and 4 GHz) at room temperature. Bottom: 3D plots of $\Delta P$ vs $H_{dc}$ curves at different frequencies ($f$ = 0.1 - 4 GHz) for samples (c) $R$ = Pr and (d) $R$ = Nd.

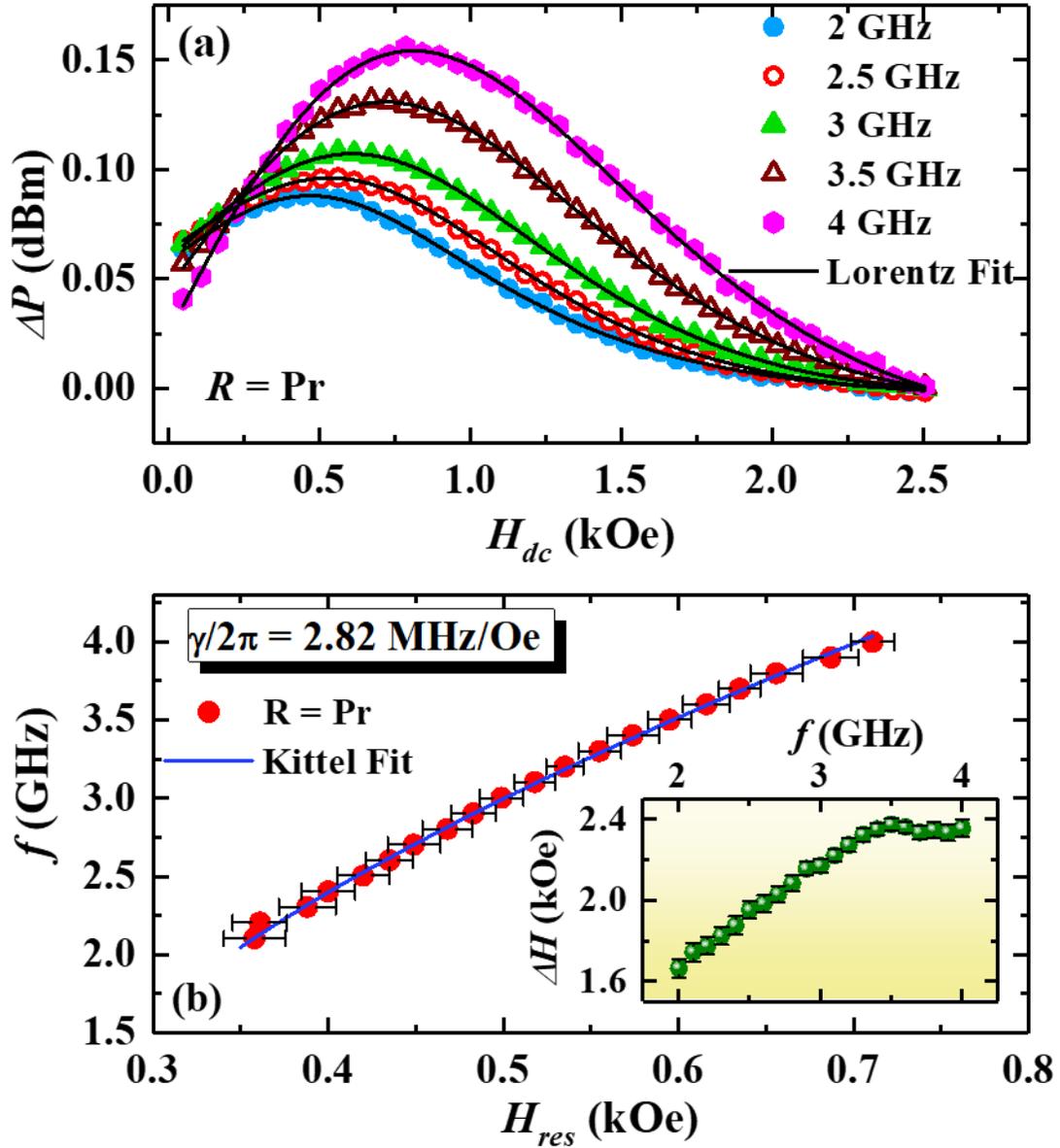

FIG. 4. A. Chanda et al.

**FIG. 4**(a) Lorentzian fit of the $\Delta P$ line shapes for the sample $R$ = Pr at selected frequencies between $f$ = 2 - 4 GHz, (b) Main panel: $f$-$H_{res}$ curve obtained from the $\Delta P$ line shape analysis for the sample $R$ = Pr with the Kittel fit and inset shows $f$ dependence of the line width $\Delta H$.

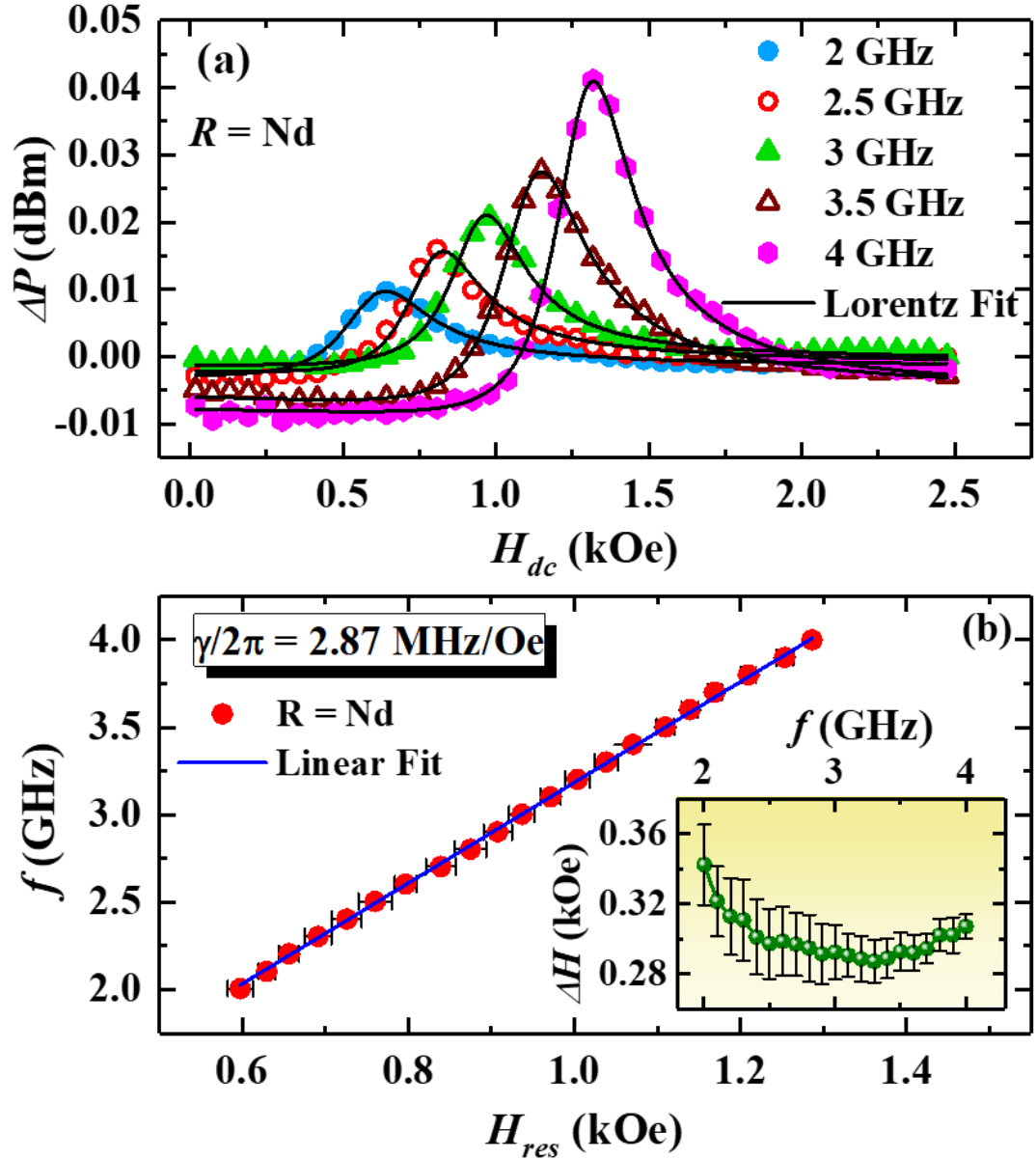

FIG. 5. A. Chanda et al.

**FIG. 5**(a) Lorentzian fit of the $\Delta P$ line shape for $R$ = Nd sample at selected frequencies between $f$ = 2 - 4 GHz, (b) Main panel: liner fit to the $f$-$H_{res}$ curve obtained from the $\Delta P$ line shape analysis and inset shows $f$ dependence of the line width $\Delta H$.

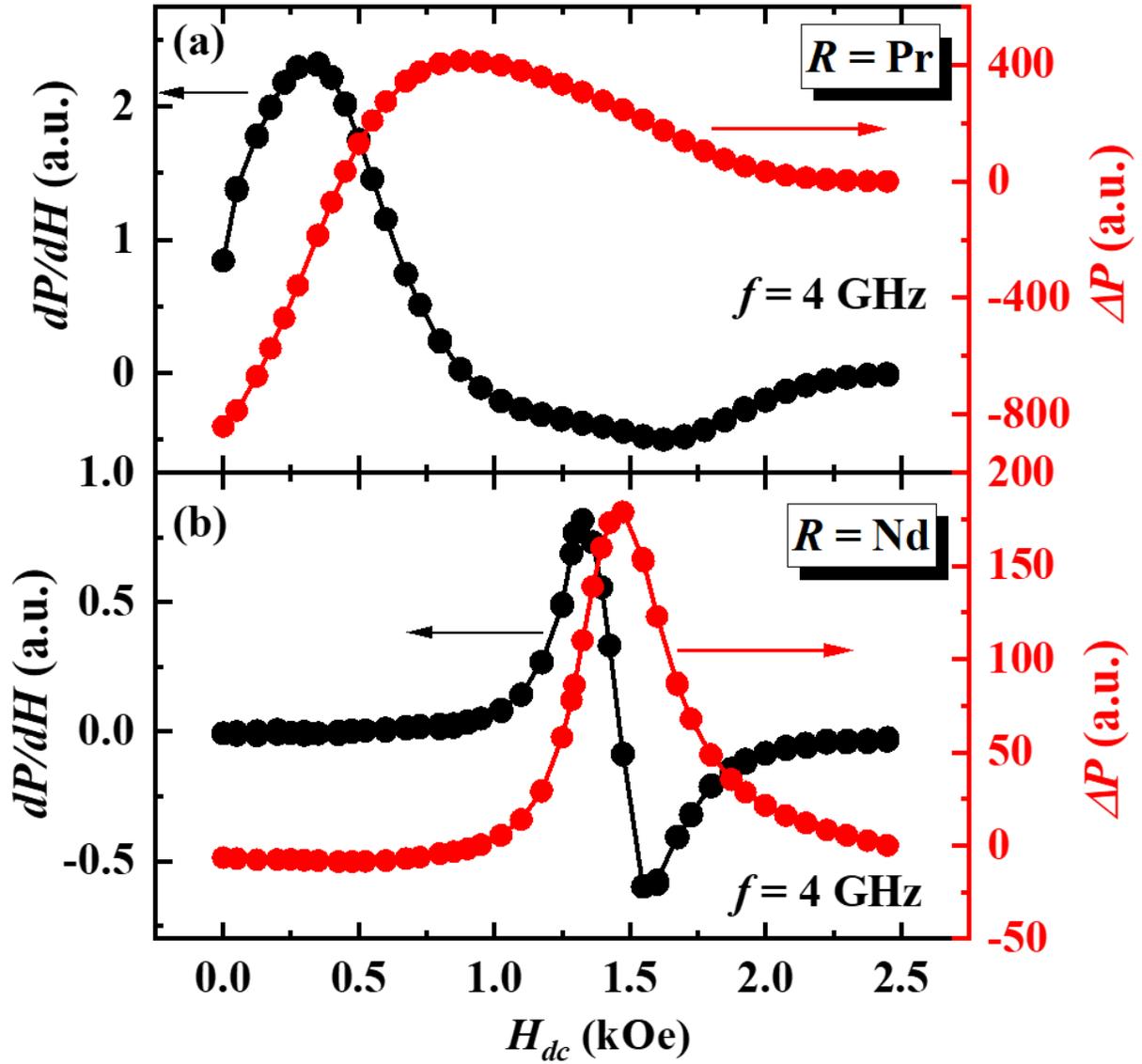

**FIG. 6** Magnetic field dependence of the field derivative of the microwave power absorption ($dP/dH$) at $f = 4$ GHz [left $y$-scale] measured using broadband FMR spectrometer and corresponding change in power absorption ($\Delta P$) estimated by integrating the $dP/dH$ data over the measured field range (right-$y$ scale) for the samples (a) $R =$ Pr and (b) $R =$ Nd.